\documentclass[aps,prd,twocolumn,showpacs,superscriptaddress,groupedaddress]{revtex4}  % for review and submission

\usepackage[colorlinks, pdfborder={0 0 0}, plainpages=false]{hyperref}
\usepackage{graphicx}
\usepackage[usenames,dvipsnames]{color}
\usepackage{amsmath, amssymb, amsfonts}
\usepackage{xspace} % Sensible space treatment at end of simple macros
\usepackage{dcolumn}% Align table columns on decimal point
\usepackage{bm}% bold math
\usepackage{multirow} % Multiple rows in tables
\usepackage{ulem}
\usepackage{breakurl}
\usepackage{enumitem}

\newcommand{\ND}{{\mathcal{N_D}}}

\newcommand{\NP}{{\mathcal{N_P}}}

\newcommand{\NT}{{\mathcal{N_T}}}
\newcommand{\SY}{{\mathcal{S}_1}}
\newcommand{\SN}{{\mathcal{S}_0}}
\newcommand{\params}{\boldsymbol \theta}
\newcommand{\nparams}{\boldsymbol \eta}
\newcommand{\LALInf}{\emph{LALInference}}
\definecolor {darkgreen}{rgb}{0.2,0.7,0.2}
\newcommand{\btheta}{\boldsymbol{\theta}}
\newcommand{\bkappa}{\boldsymbol{\kappa}}
\newcommand{\boldeta}{\boldsymbol{\eta}}

\DeclareMathOperator{\var}{Var}

\definecolor{CiteColor}{rgb}{0, 0.5, 0}
\hypersetup{citecolor=CiteColor} %
\definecolor{RefColor}{rgb}{0.55, 0, 0}
\hypersetup{linkcolor=RefColor} %

\usepackage{ulem}
\begin{document}

\title{Fortifying the characterization of binary mergers in LIGO data}

\author{\surname {Tyson} B. Littenberg}\affiliation{Center for Interdisciplinary Exploration and Research in Astrophysics 
(CIERA) \& Dept of Physics and Astronomy, Northwestern University,  2131 Tech Drive, Evanston IL, 60208, USA}
\author{\surname {Michael} Coughlin}\affiliation{Institute of Astronomy, Madingley Road, Cambridge CB3 0HA, UK}
\author{\surname {Benjamin} Farr}\affiliation{Center for Interdisciplinary Exploration and Research in Astrophysics 
(CIERA) \& Dept of Physics and Astronomy, Northwestern University,  2131 Tech Drive, Evanston IL, 60208, USA}
\author{\surname {Will} M. Farr}\affiliation{Center for Interdisciplinary Exploration and Research in Astrophysics 
(CIERA) \& Dept of Physics and Astronomy, Northwestern University,  2131 Tech Drive, Evanston IL, 60208, USA}
%\author{\surname {Benjamin} F. Farr}\affiliation{Center for Interdisciplinary Exploration and Research in Astrophysics 
%(CIERA) \& Dept of Physics and Astronomy, Northwestern University,  2131 Tech Drive, Evanston IL, 60208, USA}
%\author{\surname {Will} M. Farr}\affiliation{Center for Interdisciplinary Exploration and Research in Astrophysics 
%(CIERA) \& Dept of Physics and Astronomy, Northwestern University,  2131 Tech Drive, Evanston IL, 60208, USA}
%\author{\surname {Vassiliki} Kalogera}\affiliation{Center for Interdisciplinary Exploration and Research in Astrophysics 
%(CIERA) \& Dept of Physics and Astronomy, Northwestern University,  2131 Tech Drive, Evanston IL, 60208, USA}
\date{\today}

\begin{abstract}
The study of compact binary in-spirals and mergers with gravitational wave observatories amounts to optimizing a theoretical description of the data to best reproduce the true detector output.  While most of the research effort in gravitational wave data modeling focuses on the gravitational waveforms themselves, here we will begin to improve our model of the instrument noise by introducing parameters which allow us to determine the background instrumental power spectrum while simultaneously characterizing the astrophysical signal.  We use data from the fifth LIGO science run and simulated gravitational wave signals to demonstrate how the introduction of noise parameters results in resilience of the signal characterization to variations in an initial estimation of the noise power spectral density.  We find substantial improvement in the consistency of Bayes factor calculations when we are able to marginalize over uncertainty in the instrument noise level.
   \end{abstract}

\maketitle

\section{Introduction}
\label{sec:intro}

The cornerstone sources of gravitational radiation targeted by the LIGO/Virgo observatories are mergers of binary systems comprised of compact stellar remnants such as neutron stars (NS) and/or black holes (BH).  Signal processing for ground-based gravitational wave detectors is a daunting data mining problem as merger signals are rare events of short duration, lasting tens to hundreds of seconds, buried in years' worth of instrument noise.  Extracting astrophysical information from this mountain of data requires an elaborate analysis pipeline which in turn is dependent on accurate theoretical predictions for the signals being pursued and for the instrument noise with which those signals compete.

While noise characterization is important for any measurement, it is especially vital for GW detection because most events will contribute relatively little power to the data.  In this noise-dominated regime small changes to the instrument background result in fractionally significant changes to the relative strength of the signal (or signal-to-noise ratio) and, consequently, inferences that can be made about the system parameters.  For a detailed theoretical account of the role noise plays in the process of drawing inferences from the data, see Vallisneri 2011~\cite{Vallisneri:2011ts}.  For studies investigating the impact of real LIGO/Virgo noise on astrophysical inferences using simulated gravitational wave signals see Refs \cite{vanderSluys:2009bf,Raymond:2009cv,Vitale:2011wu}.

In Gaussian noise, the data analysis strategy for compact mergers relies on Weiner matched filtering~\cite{Thorne:1987, Allen:2001ay} which, in turn, assumes an accurate model for the instrument data.    In the matched filtering paradigm we first require gravitational wave simulations, or waveforms, $h$ which can be computed for arbitrary astrophysical parameters $\params$ over the prior volume of the search.  
Then, assuming the data $d$ are comprised of a gravitational wave signal and additive Gaussian noise $n$, a matched-filtering statistic is adopted such as the signal-to-noise ratio (SNR)  $\rho \equiv (d|h)/\sqrt{(h|h)}$ or chi-squared residual 
\begin{equation}\label{eq:chisq}
\chi^2\equiv (d-h|d-h)
\end{equation}
where we have used the standard noise weighted inner product 
\begin{equation}\label{eq:nwip}
(a|b)\equiv 2\int_0^{\infty}\frac{a(f)b(f)^*+a^*(f)b(f)}{S_n(f)} df
\end{equation}
with $S_n(f)$ representing the one-sided noise spectral density, i.e., the expectation value of the instrumental noise's Fourier power.

In the analysis of LIGO/Virgo data, the search for candidate signals is performed by convolving the simulated instrument response to a grid of pre-computed trial waveforms $h(\params)$, or templates,  with the data computing a statistic that is maximized when the template identically matches the signal, and keeping events, or ``triggers,'' which exceed a threshold for statistical significance~\cite{Abadie:2010yb,Colaboration:2011np}.  The statistic employed in GW searches is similar in principle to $\rho$, but modified using statistical studies of the data to satisfy the Neyman-Pearson criterion of minimizing the false dismissal rate for a fixed false alarm probability.   Candidate events which exceed the statistical threshold and pass consistency checks across the network are further analyzed using stochastic samplers based on the Nested Sampling~\cite{Skilling, NestedSampling, Veitch:2008ur,  Feroz:2008xx} and/or Markov chain Monte Carlo (MCMC)~\cite{Metropolis:1953am,Hastings:1970,vanderSluys:2009bf,Raymond:2009cv} algorithms to characterize the probability density function of $\params$ from which inferences about the astrophysical system are then made.  The latter portion of the data analysis procedure, generically referred to as ``parameter estimation" (PE), is the focus of the work presented here.

Parameter estimation procedures, like the search,  amount to comparing predictions for the GW data with that which was actually collected by the interferometers.  Unlike the search phase, which uses a grid of waveforms to make a point-estimate of the GW signal's parameters, PE is performed over the continuous parameter space and the goal of the sampler is to find all of the locations in the prior volume which sufficiently minimize the $\chi^2$ statistic, or maximize the likelihood 
\begin{equation}\label{eq:like}
p(d|\params)\propto e^{-\chi^2/2}
\end{equation}
which, when weighted by our prior expectations for the model parameters $p(\params)$, yields the posterior distribution function
\begin{equation}\label{eq:pdf}
p(\params|d,M)\equiv \frac{p(d|\params,M)p(\params|M)}{p(d|M)}.
\end{equation}
where $p(d|M)$ is the marginalized likelihood, or evidence, computed formally by integrating the numerator of (\ref{eq:pdf}) over the full parameter space.  
The posterior distribution function encodes our knowledge about the astrophysical system and it is from this distribution that all inferences are ultimately made.
Here and throughout this paper $p(a|b)$ denotes the conditional probability density of $a$ given $b$ so, for example, the likelihood $p(d|\params,M)$ is the probability (density) that we would measure data realization $d$ given model $M$ and parameters $\params$.
Implicit in this strategy are all of our assumptions about the data, represented by $M$, regarding both the signal (the gravitational waves themselves) and the instrument noise.  Improving any aspect of $M$ effectively improves the sensitivity of the detectors, and software is a very cost effective upgrade.  The majority of development work in this regard has focused on the numerator of (\ref{eq:nwip}) through the study of gravitational wave source modeling (e.g.,~\cite{Blanchet:2006zz,Ajith:2009bn,Buonanno:2007pf,Lindblom:2008cm}) where the goal has been to develop approximate waveforms which are both accurate, precise, and computationally efficient.  The importance of waveform development can not be overstated with key challenges remaining as we approach the advanced detector era.

\begin{figure}[htbp]
      \includegraphics[width=1\linewidth]{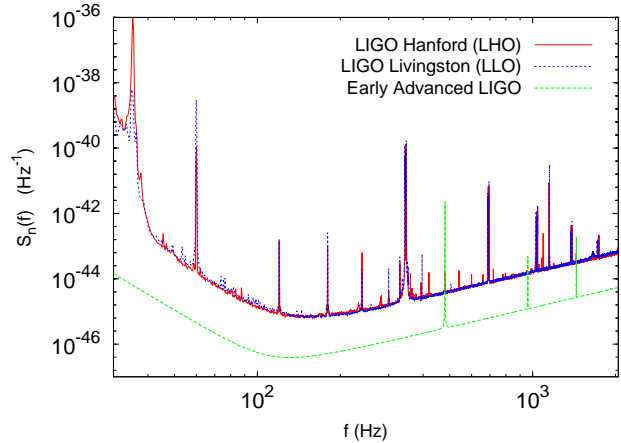} % requires the graphicx package
   \caption
   {
   	{\small Average LIGO PSDs during S5 run~\cite{PSDold}, with a prediction for the early Advanced LIGO PSD for comparison~\cite{PSD}.}
   }
   \label{fig:S5psd}
\end{figure}

Having an accurate noise model is of similar value to waveform simulations, but has historically received less attention, with the instrumental background assumed to be stable and easily characterized. The standard picture of gravitational wave detector data assumes that, in the Fourier domain representation, the noise correlation matrix is diagonal (i.e., the noise distribution is stationary~\cite{Rover:2011qd}) 
and that the noise samples are Gaussian distributed in each bin.  %Gravitational wave observatories are typically broad-band detectors but do not enjoy uniform sensitivity across the full spectrum.  Instead the noise assumes a characteristic spectral profile, the 
The GW detector noise is characterized by the one-sided power spectral density (PSD) $S_n(f) \equiv  \frac{2}{T}\langle \tilde{n}^2(f)\rangle$.  A typical additional assumption on the data model is that the PSD is both constant over the observation time, and precisely known.  Figure~\ref{fig:S5psd} shows the average PSD for LIGO from the initial design sensitivity configuration, and the expected performance of Advanced LIGO when the facilities reach full sensitivity.  %The curves can be thought of very loosely as the ``noise floor'' for the interferometers, above which signals are detectable.

The three main assumptions about the noise model -- stationarity, Gaussian distributed, and known PSD -- are overly restrictive.  The instrumental background is at best slowly varying and at worst subject to short-duration, high-amplitude, excursions of excess power, or ``glitches.''  Even in the absence of the (very problematic) glitches, failure to account for long-term changes in the noise PSD can introduce systematic errors in the signal characterization which may be comparable to the fundamental statistical measurement uncertainties.  
%While often a mild contribution to the overall error budget, 
Examples where PSD fluctuations have significantly biased the parameter estimation were encountered during the parameter estimation analysis of simulated signals added to data from the sixth LIGO and second Virgo science runs (S6 and VSR2, respectively)
%hardware injections to the extent that different, nearly contemporaneous, PSD estimates would yield disjoint posteriors for the component masses of the binary, which are a crucial ``deliverable'' from GW observation
~\cite{S6PE}. 

Parameter estimation is only half of the story, as the follow-up analyses can be employed to quantifiably assess the probability that a candidate detection is a GW event by comparing the evidence for competing models of the data.  In this context the alternative model to a GW detection is that the data contain only instrument noise.  Thus, having an adequate noise model is a prerequisite for performing meaningful evidence comparisons.% -- we can only have faith in our discoveries if we have faith in our alternatives.

Previous considerations for extending GW data analyses beyond the strict assumptions about the noise came with Allen et al in Refs.~\cite{Allen:2001ay,Allen:2002jw} to which most subsequent studies owe heritage.  The papers by Allen et al, while considering generic GW data analysis, are ultimately focused on measuring a stochastic GW background, which was more recently addressed in a space-based detector context~\cite{Adams:2010vc} where the noise must be modeled as the usual instrumental background plus a \emph{foreground} of blended gravitational-wave signals from the Galaxy.  There are several studies promoting the use of a Student's t-distribution in lieu of Gaussian noise~\cite{Rover:2008yp,Rover:2011qd}, as well as model selection recipes which employ additional coherence tests between detectors in the global interferometer network~\cite{Veitch:2009hd}.  Finally, several proof-of-principle studies using simulated data and parameterized noise models \cite{Cornish:2007if,Littenberg:2009bm,Littenberg:2010gf,Adams:2010vc} appear in the literature, building from techniques developed in response to the Mock LISA Data Challenges (e.g.,~\cite{Babak:2007zd,Babak:2009cj}).  

In this work we will employ a simple modification to the noise model by upholding our classic assumptions about the statistical properties of the noise, but adding parameters to our model of the data which fit for the PSD in each detector. 
%In doing so, our parameter estimation studies will simultaneously fit for the signal and the noise, making results robust against variations in the instrument background that are on timescales longer than the segment of data being analyzed (typically of order ten seconds).  
The additional degrees of freedom that we will introduce to the model are not strongly correlated with astrophysical parameters, thus avoiding marginalization penalties for the physical quantities of interest (e.g. mass, spin, location, etc).  
%Additionally, the PSD fitting demonstrated here opens the door to using long-term averages of the instrument noise which can, in turn, help reduce systematic errors.
%We will also investigate the impact of removing persistent spectral features (or ``lines'' -- the dramatic spikes in Fig.~\ref{fig:S5psd}) from the inner product~(\ref{eq:nwip}).  As we will show, the lines can inhibit the effectiveness of the scale parameters while contributing no useful information to the likelihood. 
The parameterized noise model demonstrated here still operates completely in the Fourier domain without modifying the underlying functional form of the noise distribution, and therefore can not account for short-duration transient and non-Gaussian noise events.  However, we find significant improvement in the consistency of the parameter estimation and model selection results when using real LIGO data.
Mitigating the effects of detector glitches, and therefore relaxing the assumptions about stationary and Gaussian noise will be left to future work where we intend to build off of the theoretical progress and simulations made in Refs.~\cite{Allen:2002jw,Rover:2011qd,Littenberg:2010gf}.
%, the inclusion of which will improve the validity of our data model and thus in the inferences drawn from the data -- particularly in our ability to distinguish between GW signals and instrument noise for events with borderline statistical significance.  It is therefore our desire to use this work not as a means to an end, but as a spring-board 
% towards implementing a more generic noise model, drawing from the most successful features of previous work \cite{Rover:2011qd,Littenberg:2010gf}.  The end-goal is to have the parameter estimation software provide useful insight to both astrophysicists \emph{and}  instrumentalists.

\section{The LIGO power spectral density}\label{sec:psd}
\begin{figure}[htbp]
      \includegraphics[width=1\linewidth]{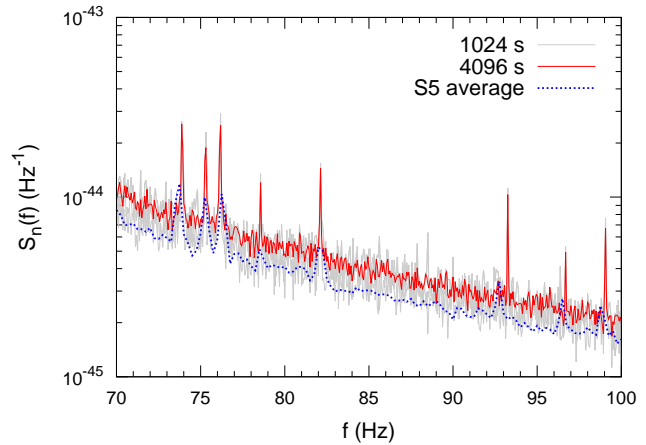} % requires the graphicx package
   \caption
   {
   	{\small Comparison of PSD estimates over a small fraction of the full LIGO bandwidth made with the prescription applied in LIGO/Virgo parameter estimation software  and using different durations of data to characterize the instrument background.  The gray line is a superposition of four consecutive 1024 s blocks of data, divided into 16 s segments for which the PSD is computed and then averaged across the block.  1024 s is typical for PE follow-up of triggers from the search pipeline.  The red (solid) line shows the average PSD if the full 4096 s of data are averaged.  The blue dotted line is the average PSD for the entire science run.}
   }
   \label{fig:psddiff}
\end{figure}

GW observatories are (nearly) continuously collecting data as all-sky monitors awaiting transient events.  Without the ability to make accurate, real-time estimates of the instrument background, previous incarnations of the parameter estimation analysis have leveraged the scarcity of GW events by using ``off source'' segments of data which are temporally ``near by'' a significant trigger produced by the matched-filtering search.  The off-source data are assumed to contain only instrument noise and are divided into smaller segments used to estimate the PSD via Welch filtering.  %The vague jargon used to describe the PSD estimation procedure -- 
The terms \emph{off source} and \emph{near by}  are not well defined or motivated, and often come down to an ad hoc procedure of looking at the data around a trigger and choosing times that do not have any obvious pathologies (glitches, highly irregular PSDs, etc.).  To date the PE procedures have had to make trade-offs between long-duration averages of the PSD for precision and shorter contemporaneous estimates of the noise level to combat slow drift of the PSD.  Short-duration data segments are not ideal for PSD estimation, as the Welch averaging is noisy in its own right.  Consequently, the prescription used in past analysis can be limited by availability of data for PSD estimation, perhaps due to gaps in the data collection or obvious glitches.  The challenges of adequately estimating the PSD from off-source data will be exacerbated in the Advanced LIGO/Virgo data because, due to the improved sensitivity at low frequencies, the amount of time when GW in-spiral signals are in the sensitive band of the detector will be increased substantially over previous incarnations of the detectors.  For low-mass sources, such as binary neutron star mergers, several hours worth of data will be necessary for PSD estimation.
Figure~\ref{fig:psddiff} shows how the goals of PSD estimation are in tension with the protocol for characterizing the instrument background.  The grey trace is the superposition of four different PSD estimates made from successive 1024 second segments of data.  The solid (red) line is the noise spectrum of the same data where the entire 4096 seconds were used for its determination.  Both example PSDs are compared to the average noise level for the entire S5 run, shown in the dashed (blue) line.  Intuitively it is clear that the random fluctuations in the noise estimate -- which effectively add noise to the measurement -- are suppressed by using more data, but at the expense of accuracy as the PSD slowly drifts. 

\begin{figure}[htbp]
      \includegraphics[width=1\linewidth]{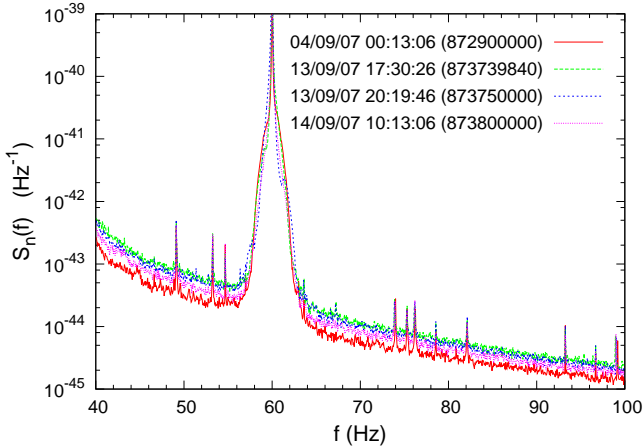} % requires the graphicx package
   \caption
   {
   	{\small Example PSDs averaged over 4096 s of initial LIGO science data from the Livingston Observatory.  Secular drift in the noise is subtle, but does significantly impact the characterization of gravitational wave signals. Dates are in dd/mm/yy format, times are in UTC, with the GPS time for the start of the data segment in parenthesis.}
   }
   \label{fig:psdflux}
\end{figure}

During times when the GW detector's noise characteristics satisfy the assumptions of Gaussianity and long-term stationarity, the outlined procedure for noise estimation is adequate for detection, characterization, and to some extent, model selection.  
%LIGO and Virgo are extraordinarily sensitive instruments, and thus inevitably experience subtle but prevalent variations to the PSD due to changing instrument behavior and environmental disturbances.  For instance, the detectors are noticeably ``quieter'' at night when there is less activity at both the observatories and the surrounding environs.  
%Despite the best-effort to get PSD estimates as up-to-date as possible for the Bayesian follow-up, 
Nonetheless, data analysis procedures must be prepared for the inevitability of discrepancies between the estimated instrument background and the actual noise in the data being analyzed.  Figure~\ref{fig:psdflux} shows an example of such variations occurring over several days using data collected when LIGO was operating at its initial design sensitivity.  The figure focuses on the lower frequencies of the sensitivity band because at high frequency the noise behavior is much more stable.  Dates are in the dd/mm/yy format, and times are in UTC, with the GPS time for the beginning of the data segment in parenthesis.  The data used to estimate the PSDs was collected at the LIGO Livingston Observatory (local time is UTC-6 hours).
%, and to bring attention to the large-scale variability common to narrow-band high-power features, or ``lines'' in the data -- the particularly dramatic feature at 60 Hz because these detectors use electricity and are located in the US.
By allowing the follow-up analysis to fit for the PSD level while simultaneously characterizing the GW signal the restrictions which require sufficiently quiescent data near a candidate event can be relaxed, instead allowing for long-term estimates of the PSD to be used.  %The parameter estimation results will now be marginalized over our uncertainty in the noise level.  
In this paper we will demonstrate such an adaptation to the model on real detector data -- a first for this type of noise modeling which has, until now, been restricted to proof-of-principal studies on simulated data.

Before delving into the supporting examples, we must first understand how the noise enters into the data analysis formalism by considering discretely sampled noise-only data from a single interferometer.  The probability of collecting $N$ noise samples $\tilde{\bf{n}}$ when noise has zero bin-to-bin correlations
% (because it is stationary)  
is  
\begin{equation}\label{eq:likelihood}
p(d) \equiv  \prod_i^N \frac{1}{\sqrt{\pi} \sigma_{i}}e^{-\frac{|\tilde{n}_{i}^2| }{\sigma_{i}^2}}
\end{equation}
where $\sigma_i^2 = \frac{1}{2}S_{n,i}$,  the PSD is $S_{n,i}$, and the summation is performed over positive frequencies only.
From this fundamental assumption about the character of the instrument noise using (\ref{eq:chisq}) and the discretized version of (\ref{eq:nwip}) it is clear how we arrive at the likelihood function (\ref{eq:like}) ubiquitously applied in gravitational wave data analysis.  
%This quasi-derivation hopefully crystallizes how the likelihood function is merely the noise model, and thus how assumptions about the noise form the backbone of the entire data analysis strategy.

To relax the assumption that we precisely know the PSD we introduce scale parameters $\boldsymbol\eta$ which act as piecewise multipliers to the PSD, modifying the likelihood as follows~\cite{Littenberg:2009bm}:
\begin{eqnarray}\label{eq:parameters}
S_{n,i} &\rightarrow& \eta_j S_{n,i},\ i_j < i \leq {i_{j+1}} \nonumber \\
\ln p(d|\params,\NP) &=& -\frac{1}{2}\left( \chi^2  + \sum_jN_j\ln\eta_j\right)+{\rm const.}
\end{eqnarray}
Each $\eta_j$ spans $N_j$ Fourier bins from $i_j$ to $i_{j+1}$.  The logarithmic term in (\ref{eq:parameters}) comes from the normalization in (\ref{eq:likelihood}) which is normally left out of any calculations because it does not depend on waveform parameters and because \emph{relative}, rather than absolute, likelihoods are employed in the PE procedures. There is an independent set of $\boldsymbol\eta$ for each detector in the network, and an additional summation over interferometers must be included in (\ref{eq:parameters}).

%Lines 

\section{Methodology}
To test the merits of the PSD-fitting noise model we use data from the end of LIGO's fifth science run (S5).  S5 began on 5 November, 2005 and was completed on 30 September, 2007, yielding a series of novel, though null, studies pertaining to limits placed on gravitational wave emission by nearby pulsars and event rates for compact binary mergers, and cosmological backgrounds~\cite{Abbott:2007rh,2008ApJ...683L..45A,Abbott:2009up,Abadie:2010yb,StochasticBackgroundS5}.  Since then, the LIGO and Virgo facilities have undergone a series of upgrades, and an additional science run (S6 and VSR2), to substantially improve the prospects of directly detecting gravitational radiation and realizing the promises of this long-anticipated new field in astronomy. 

Virgo data from VSR1 were also available during part of S5 but we have restricted this study to LIGO-only data for the following reasons:  Having only two detectors in the network puts stress on the parameter estimation methods due to strong degeneracies between extrinsic parameters (spatial location and orientation of the source) and an auxiliary goal of this work was to improve efficiency in such a network configuration.  Furthermore, it is possible that early detections will come from a two detector network as during any given science run the individual detectors have duty cycles below 100\%.  For example, $\sim75\%$ of the ``coincident time,'' defined as having more than one interferometer producing good science data, during the S6/VSR2 run had only two detectors operating~\cite{Colaboration:2011np}.  Additionally, the LIGO and Virgo facilities are operating on different schedules to complete the upgrades and commissioning of the advanced detectors.

We use the Markov chain Monte Carlo software found in the LIGO Algorithm Library (LAL) package \LALInf which was one of the samplers used for the parameter estimation studies following significant events from the S6 science run~\cite{S6PE}, to which we have included optional noise-model parameters.  Complete details of the sampler's specific implementation in \LALInf will be available in a forthcoming publication.  The MCMC algorithm implemented in LAL leans heavily on parallel tempering~\cite{Swendsen:1986} to efficiently sample the complicated posterior distribution function we have come to expect from gravitational wave detections.  In parallel tempting, several Markov chains are run simultaneously, each sampling the target distribution tempered by a  parameter $T$ such that $p(d|\params,M,T)\rightarrow p(d|\params,M)^{1/T}$.  Chains can exchange parameters while preserving detailed balance thereby improving the efficiency with which the $T=1$ chain samples the target posterior distribution function.  We further utilize the high temperature chains to compute model evidences using thermodynamic integration~\cite{Goggans:2004,Littenberg:2010gf} via:
\begin{equation}
p(d|M)=\int_0^1 \langle p(d|\params,M) \rangle_{\beta} d\beta
\end{equation}
where $\beta=1/T$ is the inverse temperature of the chain.

%Our modifications to the MCMC package center on the inclusion of parameters to model the instrument noise.  Because of our focus on the two-detector case, additional proposal distributions were introduced for waveform parameters to promote adequate mixing of the extrinsic parameters (source location and orientation) known to exhibit strong correlations, if not degeneracies, which are often decoupled by including additional detectors in the network~\cite{Rover:2006bb,Nissanke:2011ax,Veitch:2012df}.  Even though the full realization of the LIGO/Virgo collaboration is to have multiple detectors simultaneously collecting data in a truly global endeavor, we see value in having the samplers prepared for the challenges that a two-detector network presents.

The exact implementation of the PSD scale parameters can be adjusted however we did not find a strong influence on the results for the GW parameters based on different choices for the noise parameterization.  
%A larger-scale systematic study to optimize the noise model could nonetheless prove to be useful.  
The most obvious adjustable features for the PSD parameters are the number of scale factors per interferometer and the priors used for these same parameters.  We use eight scale parameters per interferometer, spaced logarithmically in frequency giving the noise fitting more fidelity at lower frequencies where the PSD fluctuations are most prevalent.  Our choice of prior in this work is the same as originally described in~\cite{Cornish:2007if,Littenberg:2009bm}, where we assume that the priors are gaussian with widths determined by the number of Fourier bins over which each scale factor is applied.  The scheme for proposing a new scale parameter $\eta_{j,y}$ from the current value $\eta_{j,x}$  employs a draw from a normal distribution with variance proportional to the width of the prior.   To wit:
\begin{eqnarray}\label{eq:etamodel}
p(\eta_j|M) &\propto& e^{-\frac{\left( 1-\eta_j\right)^2}{2\sigma_{j}^2} } \nonumber \\
\eta_{j,y} &=& \eta_{j,x} + N(0,\sigma_{j}^2/10) \nonumber \\
\sigma_j &=& \alpha \times 1/\sqrt{N_j}
\end{eqnarray}
where $N_j$ is the number of Fourier bins over which $\eta_j$ is applied to $S_n(f)$ and $\alpha$ is used to broaden the prior distribution, as $\alpha=1$ would be used for the case where we expect no long-term variation in the noise level.  For the results reported here we used $\alpha=10$ but found that the GW parameter estimation were not strongly affected by this choice for reasonable values of $\alpha$.  
The scale parameters are independent across interferometers as well as frequency blocks, so the joint prior is simply the product of $p(\eta_j|M)$. 
%In this application, we have elected to have the blocks of Fourier bins which share a common $\eta_j$ spaced evenly in $\log f$, giving the noise fitting more fidelity at lower frequencies where the PSD fluctuations are most prevalent. 

\section{Results}

To test the model's performance on real LIGO instrument noise we selected from the S5 data 300 segments with duration of 4096 seconds. As the LIGO interferometers are susceptible to noise artifacts (i.e. glitches), data quality flags identifying times of poor detector behavior were developed and used in GW searches during S5~\cite{S5DataQuality}. We have chosen our segments to fall within times that pass all of the available data quality cuts.  For each segment of data we estimated the PSD using the Welch filtering as implemented in \LALInf.  We then selected as the data to be analyzed a 16 second segment from 13 September 2007 (GPS time 873750000), and performed various tests on that data using the PSDs from throughout S5.  We will refer to the different models being tested using the following notation:
\begin{enumerate}[label=(\roman{*})]
\item Model $\ND$:  Gaussian noise with fixed PSD.
\item Model $\NP$:  Gaussian noise with variable PSD.
\item Model $\NT$:  Student's t-distributed noise with fixed PSD and degrees-of-freedom parameter $\nu$.
\end{enumerate} 
Each run of the MCMC could also be performed with (Model $\SY$) or without (Model $\SN$) including a gravitational wave template in the model.

\subsection{Noise-only model comparison} \label{sec:noiseonly}
Our first study compared the performance of the model on noise-only data, focusing on four PSDs estimated at times relatively near-by the data being analyzed (13 Aug, two segments from 12 Sept., and one from 13 Sept., all in 2007).  

\begin{figure}[htbp]
\centering
      \includegraphics[width=1\linewidth]{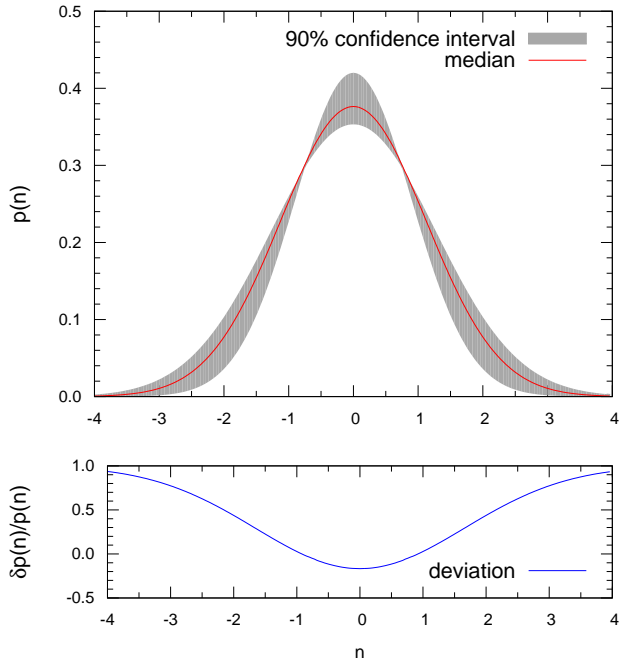} % requires the graphicx package
   \caption
   {
   	{\small Posterior distribution for the whitened noise $n\rightarrow n/\sqrt{S_n(f)}$ from 40-55 Hz.  In the top panel the line represents the median noise probability density function from an MCMC analysis, while the grey shaded region encompasses the 90\% confidence region.  The bottom panel shows the residual when the best-fit Gaussian is subtracted from the median.  We find that including PSD-fitting effectively broadens the tails of the underlying expected noise distribution.}
   }
   \label{fig:noisemodel}
\end{figure}

The top panel of figure~\ref{fig:noisemodel} shows a representative posterior for the whitened noise distribution $p(n/\sqrt{S_n(f)})$ over a frequency block with a common value for $\eta$, spanning $\sim15$ Hz and encompassing 240 data samples.  The grey shaded region spans the 90\% credible interval for the noise model and the red (solid) line shows the median noise distribution.  Effectively what we show here is the superposition of the different noise models used by the Markov chain over a 15 Hz band in the data.  By comparison, the noise distribution using the true PSD would be a unit variance, zero mean Gaussian.   Using the median distribution (red, solid curve), we compute the relative error $(p_{\rm med}(n)-p_{\rm best}(n))/p_{\rm best}(n)$, shown in the lower panel of fig.~\ref{fig:noisemodel}.   From this figure we can see how marginalizing over the PSD effectively introduces broader tails to the theoretical noise distribution, despite still using a Gaussian distribution for the noise at any point in the Markov chain.

\begin{figure}[htbp]
\centering
      \includegraphics[angle=270, width=1\linewidth]{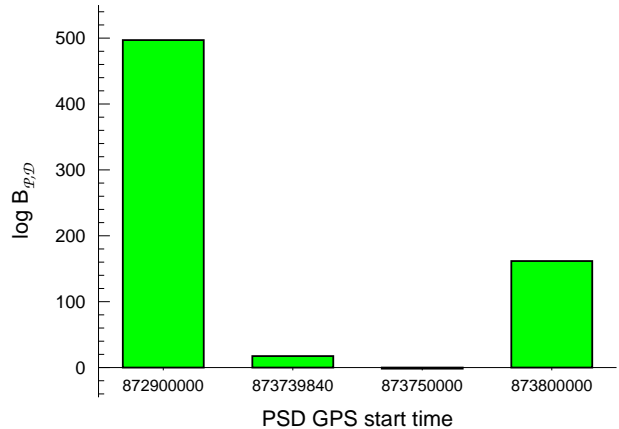} % requires the graphicx package
   \caption
   {
   	{\small Bayes factors computed for noise-only data comparing $\NP$ and $\ND$ using different PSDs  marked by the GPS time of the noise estimate.  Models which include PSD parameters are heavily favored except for 873750000, which included the data being analyzed in the noise estimation.  In this example $\ND$ is favored with 5:1 odds -- not sufficient to discriminate between models.  This implies that the PSD-fitting does no harm to the analysis when it is un-needed, and makes a substantial improvement otherwise.}
   }
   \label{fig:evidence}
\end{figure}

We repeat the analysis using four different initial PSD estimates and compute the Bayes factor between $\NP$ and $\ND$ to discern wether or not adding noise parameters sufficiently improves our fit to the data to overcome the ``Occam factor'' incurred by enlarging the prior volume of the model by introducing new degrees of freedom.  Our results are to be found in figure~\ref{fig:evidence} where the marginalized likelihood ratio for model $\mathcal{B}_{\NP,\ND}$ is shown as a function of the GPS time for the PSD estimation. 
We generally find very strong support for the PSD-fitting model ($\NP$).
The only example where the fixed-PSD model was favored, 873750000, used the data being analyzed when making the initial noise estimation.
The Bayes factor comparing $\NP$ and $\ND$ in this example resulted in 5:1 odds in favor of the fixed-PSD model since the PSD estimation in this case is accurate
%. Encouragingly, the evidence in dis-favoring the model $\NP$ is not so large that we should be concerned about ``over-fitting'' the data in circumstances where $\boldsymbol{\eta}$ are un-needed 
and does not require the additional complexity in the model.  However, from a model selection standpoint, the two models are similarly supported by the data -- a Bayes factor of 5:1 is not typically considered a strong discriminator between models.  Thus the noise-fitting model is roughly on par with the standard procedure when the initial PSD estimate is accurate.  On the other hand, if we move off-source to estimate the PSD, as is done in the real analysis, we see \emph{overwhelming} support for the PSD-fitting model (notice the y-axis in figure~\ref{fig:evidence} is logarithmic Bayes factor, so even though the support for $\NP$ in example 873739840 seems small, the odds ratio is larger than $2\times10^7:1$).  To be fair, PSD estimation is usually performed at times much closer to the candidate event but that requirement is born out of the knowledge that there \emph{is} secular evolution of the PSD and ends up restricting the amount of data that can be used to characterize the instrument noise, thereby increasing stochastic fluctuations in the PSD estimation itself.  The capability that we are demonstrating here frees the analysis of that constraint and opens the possibility of using long term, smooth, estimates for the noise.

\subsection{Statistical improvement in parameter estimation}
The true motivation of this study is not to compare noise-only models, but to see how the PSD fitting can impact the characterization of GW detections.
To that end, we now simulate a gravitational wave detection by adding a binary in-spiral waveform to the data collected at GPS time 873750000 and repeat the analysis including the signal model $\SY$.

We selected as our signal model quasi-circular, non-spinning, stationary phase waveforms computed out to the 3.5$^{\rm th}$ post-Newtonian (pN) order in the phase, taking only the leading-order amplitude -- the so-called TaylorF2 waveforms from~\cite{Buonanno:2009zt}.  Because this study is focused on the instrument noise, and not a waveform study, the TaylorF2 templates were chosen because of their computational efficiency.  We do not anticipate significantly different results for more physically detailed waveforms.  We then compare posterior distribution functions from the various data models as well as the evidence for detection in the form of the Bayes factor between models with 
%($\mathcal{B}_{\SY,\SN}=p(d|\SY,\NP)/p(d|\SN,\NP)$) 
and without 
%($\mathcal{B}_{\SY,\SN}=p(d|\SY,\ND)/p(d|\SN,\ND)$) 
noise parameters.  All injections are set to have signal-to-noise ratio between 1 and 20, depending on the particular example.

The \LALInf MCMC sampler was run on the same data using the PSDs estimated from the other 300 S5 data segments, with and without noise-fitting.  Our figure-of-merit for this study is the distribution of the medians, and the 90\% credible intervals, from each run.  We focus on the chirp mass $\mathcal{M}\equiv (m_1 m_2)^{3/5}/(m_1+m_2)^{1/5}$ because it shows the most sensitivity.  In an ideal setting the PSD would be known a priori and both the median and credible interval would be identical from run to run -- the data are not changing from one example to the next -- but long term variations in the PSD, non-stationary and/or non-Guassian features in the noise used to determine the PSD, and the finite amount of data available for noise estimation all conspire to add systematic errors to our GW measurement and thus modify the shape of the posterior distribution from one example to the next.  

\begin{figure*}[htbp]
\centering
      \includegraphics[width=1\linewidth]{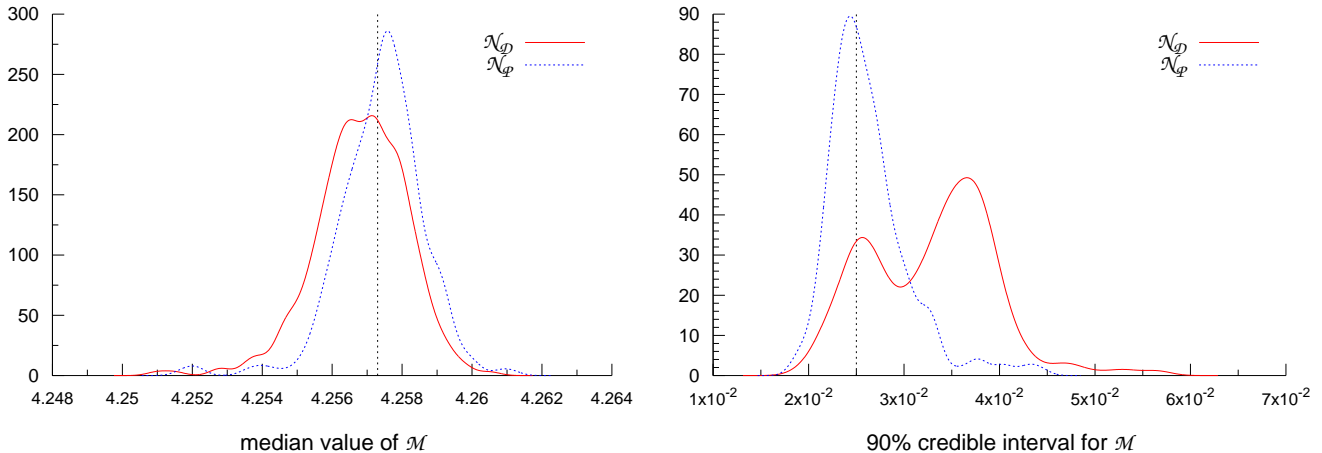} % requires the graphicx package
   \caption
   {
   	{\small Distributions of median [left] and width of 90\% credible intervals [right] obtained from the MCMC samples after analyzing the same data using 300 different initial PSD estimates.  The red (solid) histograms come from the model which assumes a perfectly-known PSD, while the blue (dotted) histogram is for a model with PSD-scale parameters.  The vertical dashed line denotes the value when the PSD estimation came from the data being analyzed (prior to the GW signal being injected).  Using the PSD scale parameters we find more consistency in the posterior distribution function across different times used to estimate the PSD.}
   }
   \label{fig:confidence}
\end{figure*}

Encouragingly, we find significant improvement in the consistency of the parameter estimation when we use the $\NP$ model.  Figure~\ref{fig:confidence} shows the distribution for the median (left panel) and the width of the 90\% credible intervals (right panel) over all 300 runs using different times to estimate the PSD.  If our PSD model were perfect there would be no difference from run to run.  In both figures the red (solid) distribution is for the fixed-PSD model ($\ND$), and the blue (dotted) histogram comes from the results of the variable-PSD model ($\NP$).  The vertical dashed grey line in each example is the value obtained when using the PSD which was estimated from the data being analyzed (prior to us adding the gravitational wave signal), and is therefore assumed to be the ``true'' value for the median and the size of the statistical error.  We find that by including PSD scale parameters the consistency from run to run for the location of the posterior distribution function (the median) and its width (the 90\% credible interval) are more narrowly peaked around the true value.  
%Despite the improvement seen here, we are not satisfied with this result.  The posteriors should be \emph{identical}, up to sampling errors from the MCMC, from run to run (the data have not changed).  Our PSD-fitting procedure is clearly a step in the right direction, but there is still room for improvement in the way we parameterize the noise model.  

\subsection{Comparison with Student's t-likelihood}
We will now take a slight detour in our study to address an alternative method for mitigating the effects of real detector noise -- using a Student's t-distribution as the underlying noise model.  This was originally advocated in a gravitational wave context by R\"{o}ver et al~\cite{Rover:2008yp} and has shown promise in studies where it has been applied to both simulated and real LIGO data~\cite{Rover:2011qd}.  In this paradigm, the likelihood becomes
\begin{equation}\label{eq:studT}
\ln p(d|\params,\NT) = -\sum_j \frac{\nu_j+1}{2}\ln\left(1+\frac{\chi_j^2}{\nu_j}  \right)+C(\nu_j)
\end{equation}
where the sum on $j$ is performed over all of the data, $\chi^2_j$ is the whitened residual power in a single bin of data, and $C(\nu_j)$ is the normalization which is computed using Gamma functions in $\nu$.  In the limit of $\nu_j\rightarrow\infty$ the Student's t-likelihood recovers the standard Gaussian likelihood in Eq. (\ref{eq:like}).
The Student's t-distribution is the ideal theoretical choice for the likelihood when the noise is stationary and the PSD is unknown and must be estimated by a finite amount of data.  Under such circumstances the degrees-of-freedom parameter $\nu_j$ is, in principal, a constant over $j$ determined by the number of segments of data used to make the initial estimate of the PSD.  However, small values of $\nu_j$ substantially broaden the tails of the theoretical noise distribution, making the noise model robust against non-Gaussianity in the data.  R\"{o}ver found $\nu\sim10$ to work well for real detector data despite using $\sim30$ data segments to estimate the PSD.  

\begin{figure*}[htbp]
\centering
      \includegraphics[width=1\linewidth]{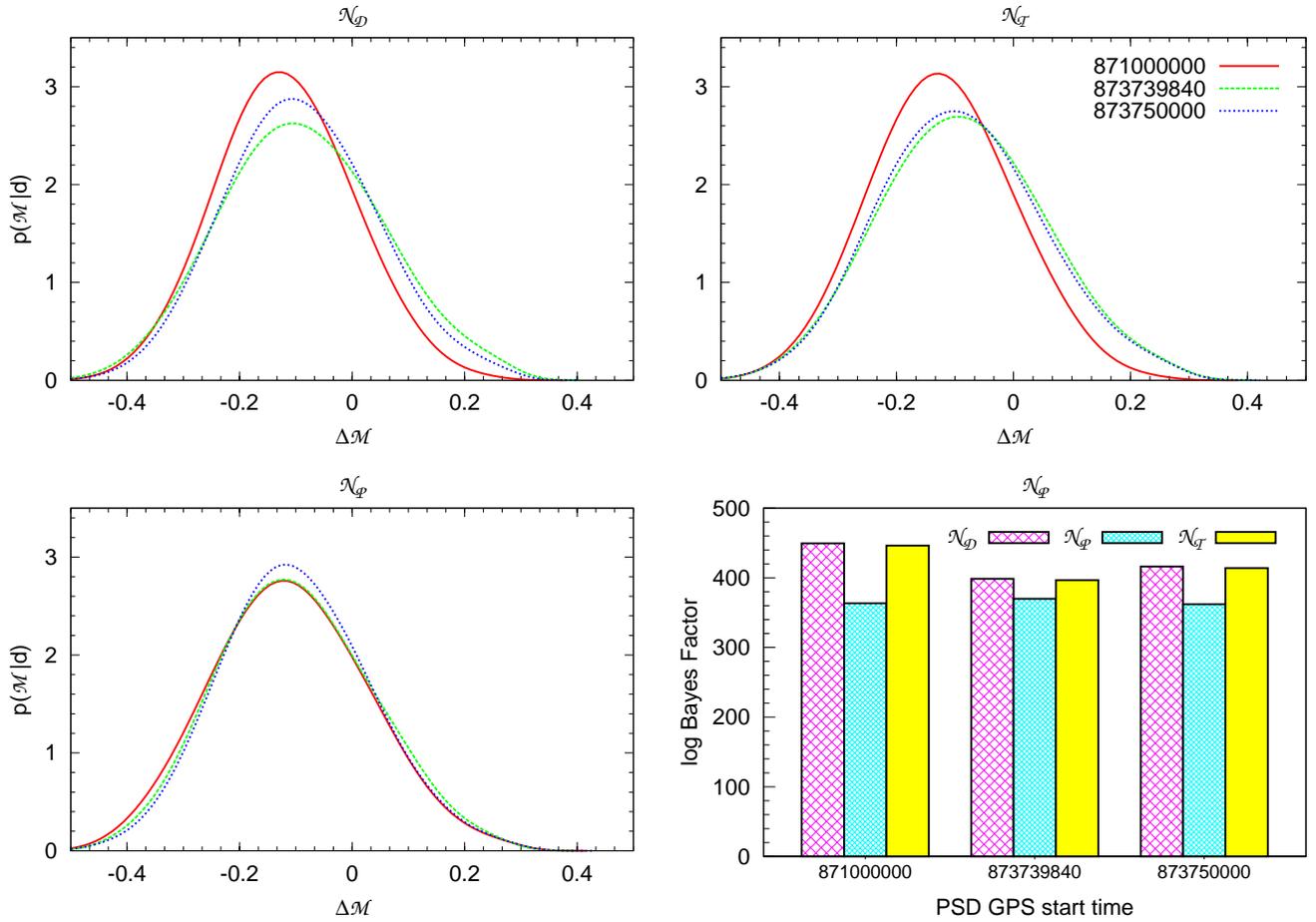} % requires the graphicx package
   \caption
   {
   	{\small Posteriors for $\mathcal{M}$ for different PSDs and models.  The top-left panel are the results of the standard version of the MCMC sampler.  The top-right panel uses one implementation of the Student's t-likelihood.  The bottom-left panel uses the PSD-fitting model.  In these examples the PSD-fitting model is the least sensitive to different PSD estimates.  The bottom-right panel shows the Bayes factor for detection as a function of PSD estimation.  In these examples the PSD-fitting model $\NP$ yields the most stable evidence ratios.}
   }
   \label{fig:mcposterior}
\end{figure*}

We ran a few of our cases using the Student's t-distribution $\NT$ to compare against models $\ND$ and $\NP$.  Results are shown in figure~\ref{fig:mcposterior} where we have again chosen to focus on the marginalized posterior distribution function for the chirp mass $\mathcal{M}$, as it shows the largest dependence to the PSD time for this example.  Notice $\mathcal{M}$ in this example is larger than before -- we used a higher-mass injection to put the ``loudest'' part of the waveform in the data where PSD fluctuations were most noticeable.  The top left panel shows the results using the usual Gaussian likelihood with no PSD-scale parameters $\ND$.  The independent variable is the displacement away from the injected value.  Different histograms correspond to the GPS time when the PSD was estimated.  Time 873750000 (blue, dashed) contains the data segment being analyzed.  The differences between chirp-mass distributions my not seem dramatic, but given that the data are the same for each run, we would like to mitigate any differences due to the PSD estimation as much as possible.  If we redo the analysis but now include the PSD fitting parameters $\NP$ (bottom left panel) we find improved consistency.  The top right panel uses no noise parameters, but adopts a Student's t-distribution with constant $\nu_j$ determined by the number of data segments used to estimate the PSD. In this example we find that the Student's t-likelihood does improve the convergence for one example (873739840, green dashed line), but could not similarly improve results using the PSD from GPS time 871000000 (red, solid line).  This does not imply that an optimized application of the Student's t-distribution likelihood, by choosing fewer degrees of freedom $\nu$, is not capable of improving this example.  The point of this demonstration is to show that differences in PE results are not simply attributed to sampling errors in the PSD estimation (which the Student's-t distribution should perfectly accommodate) and the PSD scale parameters can readily ameliorate differences between the data being analyzed and the estimated noise level.  
%A draw-back to using the Student's t-likelihood is that it is not sufficiently flexible to accommodate the large-scale drift in the PSD, in that the distribution is only able to \emph{broaden} the noise distribution, and can not account for the possibility that $S_n(f)$ as determined from near-by data may \emph{over-estimate} the true detector PSD at the time of a GW candidate.  The PSD scale parameters, on the other hand, offer much more freedom to account for slow secular drift in the PSD, only being confined by our choice of prior, but still rely on Gaussian distributed noise (apart from the effect of marginalizing over Gaussian noise distributions, as described in \S \ref{sec:noiseonly}).  

The bottom-right panel of figure~\ref{fig:mcposterior} shows an arguably more significant divergence between the standard likelihood, the Student's t, and the PSD scale parameters.  Here we show how the Bayes factor we would compute to determine the odds that a GW has been detected depends strongly on the PSD estimation.  In this example the PSD-fitting model $\NP$ (cyan, dotted) shows the least variability in the detection confidence based on the PSD estimation, while both the fixed PSD model $\ND$ (magenta, cross-hashed) and Student's t-likelihood $\NT$ (yellow, solid) exhibit similar sensitivity to the noise estimation, despite $\NT$ being able to ``correct'' the posterior for GPS time 873739840.  The Bayes factors in this example are sufficiently large ($\sim e^{400}:1$) that these fluctuations would not impact any conclusions drawn from the data.  However we will now investigate the role  $\nparams$ can play in model selection for marginal detections, when precision and accuracy in the Bayes factor computation will be paramount in evaluating candidate GW events. 

We would be remiss if we failed to reemphasize that the Student's t-distribution in these examples uses the strict interpretation of the degrees of freedom $\nu$, and that by artificially setting that to a lower value, or marginalizing over it in the way we treat the PSD scale parameters, we may temper the differences between the models.  The PSD fitting method advocated in this work is in fact closely related to the Student's t-distribution likelihood in \cite{Rover:2008yp,Rover:2011qd}.  By replacing our Gaussian prior on $\boldeta$ with an inverse-chi-squared distribution, and allowing each Fourier bin an independent $\eta$ we can analytically marginalize over the noise parameters and arrive at the Student's t-distribution for the likelihood.  If we were to adopt the inverse-chi-squared prior for the piecewise PSD scale parameters and integrate the posterior over $\boldeta$ we arrive at a likelihood that is qualitatively similar to the Student's t-distribution but maintains the broad flexibility needed to handle large-scale drift in the LIGO/Virgo PSD.  The derivation and discussion of the analytic marginalization for our PSD-fitting model can be found in the Appendix.

\subsection{Discerning betweens signal and noise models}
\begin{figure*}[htbp]
      \includegraphics[angle=270,width=1\linewidth]{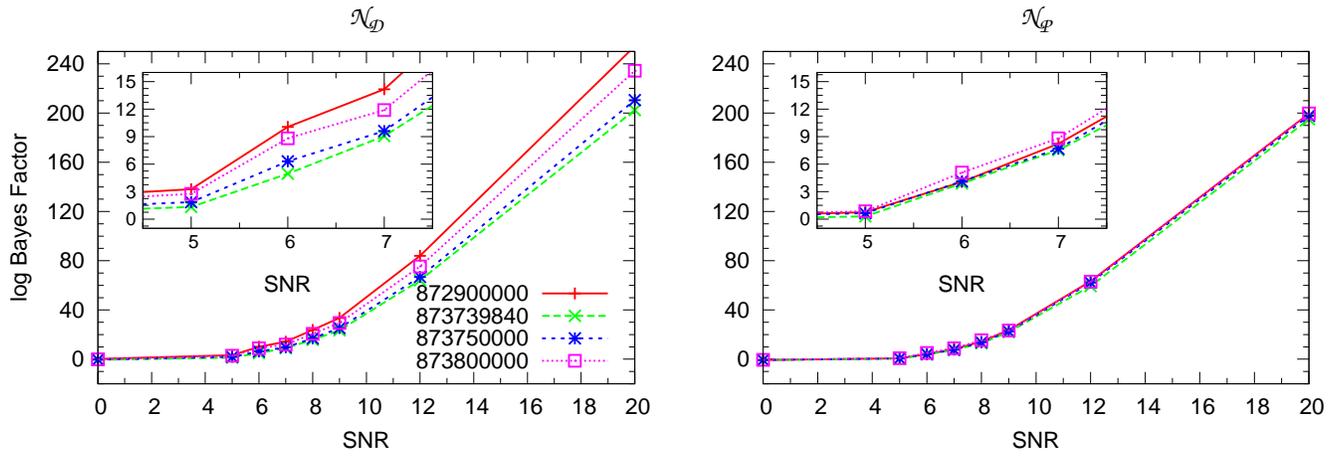} % requires the graphicx package
   \caption
   {
   	{\small Bayes factors as a function of SNR for different noise models and PSDs.  The left-hand panel shows substantial variation depending on PSD estimation.  Bayes factors that show this much variability are of limited use.  The right-hand panel shows significant improvement gained by including the PSD-fitting parameters.}
   }
   \label{fig:bayes}
\end{figure*}

For our last study, we will use the same GW system and S5 data as before, but now change the distance to the source to get a desired signal-to-noise ratio (SNR).  In doing so, we can monitor how the Bayes factor changes as we increase the signal strength, and compare our two Gaussian-noise models $\ND$ and $\NP$ with regards to their model selection capabilities.  The Bayes factor vs. SNR plots can be found in figure~\ref{fig:bayes}.  Here we see substantial variation in the Bayes factors when using noise model $\ND$ (left-hand panel), to the point where different conclusions about low SNR gravitational wave detections (see inset of left-hand panel) could potentially be made.  For example, the SNR $=6$ injection produced Bayes factors between $\sim50:1$ and $\sim2\times10^5:1$ under model $\ND$.  Bayesian model selection will not be a useful tool for discriminating between GW detection and noise if it is subject to such variability.  On the other hand, our model selection calculation is substantially more stable when we include the PSD scale parameters (right-hand panel).
%, and it is likely that more improvements are possible as we continue to hone the model and build in flexibility to accommodate non-stationary, non-Gaussian noise excursions.

\label{sec:examples}

\section{Summary and Future Work}
\label{sec:summary}
We have demonstrated, using LIGO data from the fifth science run, that adding scale parameters which modify the PSD make inferences drawn from the data more robust against slowly varying instrument noise.  The added parameters effectively allow us to marginalize our results over uncertainty in the PSD.  This work comprehensively demonstrates the importance, and the benefits, of modeling detector noise in the context of parameter estimation and model selection.

We show in figure~\ref{fig:evidence} that the noise model including scale parameters $\boldsymbol{\eta}$ receives overwhelming support when the PSD estimation comes from times separate from the data being analyzed, but garners comparable support to the fixed PSD noise model when the instrument background is estimated from the data itself (which is, of course, not possible when a potential GW event is contained in the data).

We show in figure~\ref{fig:confidence} that parameter estimation -- in the way of a point estimate (the median) and the width of the posterior distribution -- is sensitive to the PSD estimation, and that sensitivity can be significantly reduced by marginalizing over the noise level as prescribed here.  

We performed a small comparison between the PSD-fitting model and the Student's t-likelihood in figure~\ref{fig:mcposterior} and find the Student's t-distribution, as we employed it here, was not as effective as our parameterized PSD.  That being said, we see potential in combining the assets of these different approaches to solving the same problem, and anticipate a hybrid marginalized-PSD/Student's t-likelihood may prove optimal and we will pursue this and other strategies for noise modeling as we continue to improve the LIGO/Virgo noise model, such as using the analytically marginalized likelihood in the Appendix and marginalizing over hyper-parameters characterizing the prior.

Arguably the most important impact of the noise estimation is in computing the Bayes factor to select between models suggesting a GW detection, and those favoring instrument noise.  Figure~\ref{fig:bayes} shows strong dependence to the PSD estimate -- upwards of ~20\% to 50\% variation in $\log {\mathcal B}_{\SY,\SN}$ -- which is substantially reduced when the PSD fitting parameters are included.  This improvement does not appear to come at the expense of diminished sensitivity to the GW signals (which is always a risk when the dimension of a model is increased).

The noise model promoted here frees the parameter estimation follow-up analysis from its requirement to find data near to a candidate event to estimate the noise level.  Such a requirement can impede the PE results because data-quality shortcomings near a GW trigger can restrict the duration of data available for the noise estimation.  Finding enough data to estimate the PSD will become increasingly problematic in the advanced detector era as low mass binary in-spiral signals will be $\sim1000$ seconds in duration, requiring several hours of data for adequate Welch filtering -- a \emph{very} long time to hope for stationary noise. The ability to use long-term estimates of the instrument noise relaxes that restriction, mitigates the impact of transient noise events in the PSD estimation, and reduces random fluctuations introduced in the residual by the finite duration of data used to estimate the PSD.  
%We will investigate how much flexibility the noise model we have implemented here truly has, to try and anticipate at what intervals (if at all), and for what duration, we will require new PSD estimates from the data itself.

Our demonstrations using real instrument data in this paper show great promise in making  parameter estimation results more robust in the presence of real instrument noise but they are not perfect. 
% However, we are not prepared to declare victory over LIGO/Virgo noise -- this paper is not the last word on noise modeling for ground-based detectors.  
%The noise model implemented here is rather crude -- simply a piece-wise re-scaling of the measured PSD -- and could certainly be improved both by optimizing the model used here, either by adjusting the number of scale parameters, modifying the priors on $\boldsymbol\eta$, and/or changing the window sizes over which the noise is re-scaled.  
Alternative functional forms for the PSD scaling parameters, such as using parameterized power laws~\cite{Adams:2010vc} or a physical model for the noise power spectral density, may further improve noise modeling while ultimately liberating the analysis from using data outside of that which is being analyzed to provide information about the instrument background.  
%Such details need to be fully explored and optimized in a much larger-scale study before the PSD-fitting advocated here can be integrated into the mainstream parameter estimation software used on future LIGO and Virgo detections.  
%Such a study is underway and will be completed well before the first advanced era data are being collected.

While mitigating the impact of the initial PSD estimation is certainly an important technology for gravitational wave data analysis, it does not address the more fundamental assumptions about the instrument noise -- that the noise is Gaussian distributed and stationary over the duration of the data being analyzed.  This study is our first foray into applying more sophisticated noise modeling techniques into real detector data, and we are bolstered by the results here to pursue alternative, ``heavy-tailed'' distributions for the noise such as the Student's t~\cite{Rover:2011qd} or a sum of two Gaussians~\cite{Littenberg:2010gf}, as well as incorporating Bayesian model selection on glitches into the \LALInf ~software as demonstrated by the {\it BayesWave} algorithm first introduced in~\cite{Littenberg:2010gf}.

\section{Acknowledgments}
We would like to thank Nelson Christensen for thorough comments on this manuscript specifically in regards to our use of the Student's t-distribution, and Neil Cornish for suggesting the inclusion of Figure~\ref{fig:noisemodel} as a way to understand the effect of marginalizing over the PSD.  We also Vicky Kalogera, and Ilya Mandel for discussions and suggestions related to this work.  T.L. and B.F. are supported by NSF LIGO grant, award  PHY-0969820.  B.F. receives additional support from a NSF graduate research fellowship, award DGE-0824162. M.C.'s work is funded by the Winston Churchill Foundation of the United States of America.  The MCMC runs were carried out using computing resources from 
%Quest, 
the Extreme Science and Engineering Discovery Environment (XSEDE), which is supported by National Science Foundation grant number OCI-1053575, and the Nemo cluster supported by National Science Foundation awards PHY-0923409 and PHY-0600953 to UW-Milwaukee.  T.L. would like to thank the The Center for Space Plasma and Aeronomic Research (CSPAR) at The University of Alabama in Huntsville for their hospitality.

\appendix 

\section{Analytic Marginalization Over $\eta$}\label{appendix}

It is possible to analytically marginalize over the noise parameters
that appear in Eq.~\ref{eq:parameters} if a slightly different prior than
Eq.~\ref{eq:etamodel} is used.  Consider the likelihood in
Eq.~\ref{eq:parameters}:
\begin{multline}
  p(d | \btheta, \NP) = \\ \prod_j \prod_{i_j < i \leq i_{j+1}} \frac{1}{\sqrt{2\pi \eta_j S_{n,i}}} \exp\left[ - \frac{\sum_{i_j < i \leq i_{j+1}} \chi^2_i}{2 \eta_j} \right],
\end{multline}
where 
\begin{equation}
  \chi^2_i \equiv \frac{\left| \tilde{n}_i \right|^2}{S_{n,i}}.
\end{equation}
\begin{widetext}
If we impose a prior on each $\eta_j$ of the form\footnote{This is a
  scaled inverse-chi-squared distribution for $\eta_j$.}
\begin{equation}
  \label{eq:analytic-eta-prior}
  p(\eta_j | \kappa_j, \NP) = \frac{(1+\kappa_j)^{1+\kappa_j}}{\Gamma(1+\kappa_j)} \eta_j^{-(\kappa_j+3)} \exp\left[ -\frac{1+\kappa_j}{\eta_j} \right],
\end{equation}
where $\kappa_j > 0$ is a hyperparameter, then we can integrate the
dependence on $\eta_j$ out of the posterior (i.e.\ marginalize over
the $\eta_j$), obtaining
\begin{multline}
  \label{eq:marginalized-likelihood}
  p(d | \btheta', \bkappa, \NP) \equiv \int d \boldeta\, p(d | \btheta, \NP) p(\boldeta | \bkappa, \NP) \\ = \prod_j \frac{\Gamma\left( \frac{N_j}{2} + 2 + \kappa_j\right)}{\Gamma\left(1+\kappa_j\right)} \left( 1 + \kappa_j \right)^{-\left( \frac{N_j}{2} + 1 \right)} \left[\prod_{i_j < i \leq i_{j+1}} \frac{1}{\sqrt{2 \pi S_{n,i}}} \right] \left( 1 + \frac{1}{2\left( 1 + \kappa_j \right)} \sum_{i_j < i \leq i_{j+1}} \chi^2_i \right)^{-\left(\frac{N_j}{2} + \kappa_j + 2\right)},
\end{multline}
where $\btheta' = \btheta \setminus \boldeta$ are the non-noise
parameters, and recall that $N_j = i_{j+1} - i_j$ is the number of
frequency bins in the $j$th group.
\end{widetext}

The prior in Eq.~\eqref{eq:analytic-eta-prior} is chosen so that the
\emph{prior} mean of $\boldeta$ is $\textbf{1}$, and the variance is 
\begin{equation}
  \var_\mathrm{prior}(\boldeta) = \frac{1}{\bkappa}
\end{equation}
The distribution in Eq.~\eqref{eq:analytic-eta-prior} is not a
Gaussian, but, provided that $\kappa_j \ll N_j$ the prior does not
have a strong influence on the posterior.  If, for example, we choose
$\kappa_j$ to match the variance in Eq.~\ref{eq:etamodel}, that is,
\begin{equation}
  \kappa_j = \frac{N_j}{\alpha^2},
\end{equation}
with $\alpha \simeq 10$, then we do not expect that the difference
between the Gaussian prior and the inverse-chi-squared prior to matter
in the posterior.  Intuitively, as long as the prior allows for
relative fluctuations in the noise level that are greater than
$1/\sqrt{N_j}$, the uncertainty in the posterior from making 
$N_j$ independent measurements of the noise level in the frequency
bins from $i_j$ to $i_{j+1}$ will be much smaller than the uncertainty
in the prior, and the precise shape of the prior will not matter.

Note that the likelihood in Eq.~\eqref{eq:marginalized-likelihood}
differs in general from the Student's t likelihood discussed in
\cite{Rover:2011qd} and used elsewhere in this paper.  In that work,
each frequency bin is allowed to fluctuate independently in noise
level, and the prior on the noise is not constrained to have a mean of
1.  The independence of the noise bins means that there is, in effect,
one $\eta$ parameter per bin in that work (see Eq.~\ref{eq:studT}); due
to the large number of noise paremeters, measurements of the noise are
correspondingly weak, and the marginalized likelihood is much more
dependent on choice of prior.  For the choirce $\nu_j \simeq 10$, the
prior variance on the noise (which is approximately the posterior
variance on the noise, since there is only one bin per noise
parameter) is about 10\%; in the method discussed in this Appendix and
throughout the paper, the posterior variance is controlled mostly by
the $N_j$ independent measurements of the noise level associated with
one $\eta_j$ parameter, not the wide prior.  This method can therefore
cope better with the relatively large noise variations from time to
time (see Figure~\ref{fig:psddiff}) without large uncertainty on the noise
estimate from the data.

\bibliography{papers}

\begin{thebibliography}{42}
\expandafter\ifx\csname natexlab\endcsname\relax\def\natexlab#1{#1}\fi
\expandafter\ifx\csname bibnamefont\endcsname\relax
  \def\bibnamefont#1{#1}\fi
\expandafter\ifx\csname bibfnamefont\endcsname\relax
  \def\bibfnamefont#1{#1}\fi
\expandafter\ifx\csname citenamefont\endcsname\relax
  \def\citenamefont#1{#1}\fi
\expandafter\ifx\csname url\endcsname\relax
  \def\url#1{\texttt{#1}}\fi
\expandafter\ifx\csname urlprefix\endcsname\relax\def\urlprefix{URL }\fi
\providecommand{\bibinfo}[2]{#2}
\providecommand{\eprint}[2][]{\url{#2}}

\bibitem[{\citenamefont{Vallisneri}(2011)}]{Vallisneri:2011ts}
\bibinfo{author}{\bibfnamefont{M.}~\bibnamefont{Vallisneri}},
  \bibinfo{journal}{Phys. Rev. Lett.} \textbf{\bibinfo{volume}{107}},
  \bibinfo{pages}{191104} (\bibinfo{year}{2011}), \eprint{{arXiv:1108.1158
  [gr-qc]}}.

\bibitem[{\citenamefont{van~der Sluys et~al.}(2009)\citenamefont{van~der Sluys,
  Mandel, Raymond, Kalogera, R\"{o}ver et~al.}}]{vanderSluys:2009bf}
\bibinfo{author}{\bibfnamefont{M.}~\bibnamefont{van~der Sluys}},
  \bibinfo{author}{\bibfnamefont{I.}~\bibnamefont{Mandel}},
  \bibinfo{author}{\bibfnamefont{V.}~\bibnamefont{Raymond}},
  \bibinfo{author}{\bibfnamefont{V.}~\bibnamefont{Kalogera}},
  \bibinfo{author}{\bibfnamefont{C.}~\bibnamefont{R\"{o}ver}},
  \bibnamefont{et~al.}, \bibinfo{journal}{Class. Quantum Grav.}
  \textbf{\bibinfo{volume}{26}}, \bibinfo{pages}{204010}
  (\bibinfo{year}{2009}), \eprint{{arXiv}:0905.1323}.

\bibitem[{\citenamefont{Raymond et~al.}(2010)\citenamefont{Raymond, van~der
  Sluys, Mandel, Kalogera, R\"{o}ver et~al.}}]{Raymond:2009cv}
\bibinfo{author}{\bibfnamefont{V.}~\bibnamefont{Raymond}},
  \bibinfo{author}{\bibfnamefont{M.}~\bibnamefont{van~der Sluys}},
  \bibinfo{author}{\bibfnamefont{I.}~\bibnamefont{Mandel}},
  \bibinfo{author}{\bibfnamefont{V.}~\bibnamefont{Kalogera}},
  \bibinfo{author}{\bibfnamefont{C.}~\bibnamefont{R\"{o}ver}},
  \bibnamefont{et~al.}, \bibinfo{journal}{Class. Quantum Grav.}
  \textbf{\bibinfo{volume}{27}}, \bibinfo{pages}{114009}
  (\bibinfo{year}{2010}), \eprint{{arXiv:0912.3746 [gr-qc]}}.

\bibitem[{\citenamefont{Vitale et~al.}(2012)\citenamefont{Vitale, Del~Pozzo,
  Li, Van Den~Broeck, Mandel et~al.}}]{Vitale:2011wu}
\bibinfo{author}{\bibfnamefont{S.}~\bibnamefont{Vitale}},
  \bibinfo{author}{\bibfnamefont{W.}~\bibnamefont{Del~Pozzo}},
  \bibinfo{author}{\bibfnamefont{T.~G.} \bibnamefont{Li}},
  \bibinfo{author}{\bibfnamefont{C.}~\bibnamefont{Van Den~Broeck}},
  \bibinfo{author}{\bibfnamefont{I.}~\bibnamefont{Mandel}},
  \bibnamefont{et~al.}, \bibinfo{journal}{Phys. Rev. D}
  \textbf{\bibinfo{volume}{85}}, \bibinfo{pages}{064034}
  (\bibinfo{year}{2012}), \eprint{{arXiv:1111.3044 [gr-qc]}}.

\bibitem[{\citenamefont{Thorne}(1987)}]{Thorne:1987}
\bibinfo{author}{\bibfnamefont{K.~S.} \bibnamefont{Thorne}},
  \emph{\bibinfo{title}{Gravitational Radiation in 300 Years of Gravitation}},
  \bibinfo{number}{pp 330-458} (\bibinfo{publisher}{Cambridge University
  Press}, \bibinfo{year}{1987}).

\bibitem[{\citenamefont{Allen et~al.}(2002)\citenamefont{Allen, Creighton,
  Flanagan, and Romano}}]{Allen:2001ay}
\bibinfo{author}{\bibfnamefont{B.}~\bibnamefont{Allen}},
  \bibinfo{author}{\bibfnamefont{J.~D.~E.} \bibnamefont{Creighton}},
  \bibinfo{author}{\bibfnamefont{E.~E.} \bibnamefont{Flanagan}},
  \bibnamefont{and} \bibinfo{author}{\bibfnamefont{J.~D.}
  \bibnamefont{Romano}}, \bibinfo{journal}{Phys. Rev. D}
  \textbf{\bibinfo{volume}{65}}, \bibinfo{pages}{122002}
  (\bibinfo{year}{2002}), \eprint{{arXiv}:gr-qc/0105100}.

\bibitem[{\citenamefont{Abadie et~al.}(2010)}]{Abadie:2010yb}
\bibinfo{author}{\bibfnamefont{J.}~\bibnamefont{Abadie}} \bibnamefont{et~al.}
  (\bibinfo{collaboration}{LIGO Scientific Collaboration, Virgo
  Collaboration}), \bibinfo{journal}{Phys. Rev. D}
  \textbf{\bibinfo{volume}{82}}, \bibinfo{pages}{102001}
  (\bibinfo{year}{2010}), \eprint{{arXiv:1005.4655 [gr-qc]}}.

\bibitem[{\citenamefont{Abadie et~al.}(2012)}]{Colaboration:2011np}
\bibinfo{author}{\bibfnamefont{J.}~\bibnamefont{Abadie}} \bibnamefont{et~al.}
  (\bibinfo{collaboration}{LIGO Collaboration, Virgo Collaboration}),
  \bibinfo{journal}{Phys. Rev. D} \textbf{\bibinfo{volume}{85}},
  \bibinfo{pages}{082002} (\bibinfo{year}{2012}), \eprint{arXiv: 1111.7314
  [gr-qc]}.

\bibitem[{\citenamefont{{Skilling}}(2004)}]{Skilling}
\bibinfo{author}{\bibfnamefont{J.}~\bibnamefont{{Skilling}}}, in
  \emph{\bibinfo{booktitle}{American Institute of Physics Conference Series}},
  edited by \bibinfo{editor}{\bibfnamefont{R.}~\bibnamefont{{Fischer}}},
  \bibinfo{editor}{\bibfnamefont{R.}~\bibnamefont{{Preuss}}}, \bibnamefont{and}
  \bibinfo{editor}{\bibfnamefont{U.~V.} \bibnamefont{{Toussaint}}}
  (\bibinfo{year}{2004}), vol. \bibinfo{volume}{735} of
  \emph{\bibinfo{series}{American Institute of Physics Conference Series}}, pp.
  \bibinfo{pages}{395--405}.

\bibitem[{Nes()}]{NestedSampling}
\bibinfo{howpublished}{\url{http://www.inference.phy.cam.ac.uk/bayesys/}}.

\bibitem[{\citenamefont{Veitch and Vecchio}(2008)}]{Veitch:2008ur}
\bibinfo{author}{\bibfnamefont{J.}~\bibnamefont{Veitch}} \bibnamefont{and}
  \bibinfo{author}{\bibfnamefont{A.}~\bibnamefont{Vecchio}},
  \bibinfo{journal}{Phys. Rev. D} \textbf{\bibinfo{volume}{78}},
  \bibinfo{pages}{022001} (\bibinfo{year}{2008}), \eprint{arXiv:0801.4313
  [gr-qc]}.

\bibitem[{\citenamefont{Feroz et~al.}(2008)\citenamefont{Feroz, Hobson, and
  Bridges}}]{Feroz:2008xx}
\bibinfo{author}{\bibfnamefont{F.}~\bibnamefont{Feroz}},
  \bibinfo{author}{\bibfnamefont{M.~P.} \bibnamefont{Hobson}},
  \bibnamefont{and} \bibinfo{author}{\bibfnamefont{M.}~\bibnamefont{Bridges}}
  (\bibinfo{year}{2008}), \eprint{{arXiv}:0809.3437}.

\bibitem[{\citenamefont{Metropolis et~al.}(1953)\citenamefont{Metropolis,
  Rosenbluth, M.N., Teller, and Teller}}]{Metropolis:1953am}
\bibinfo{author}{\bibfnamefont{N.}~\bibnamefont{Metropolis}},
  \bibinfo{author}{\bibfnamefont{A.}~\bibnamefont{Rosenbluth}},
  \bibinfo{author}{\bibfnamefont{R.}~\bibnamefont{M.N.}},
  \bibinfo{author}{\bibfnamefont{A.}~\bibnamefont{Teller}}, \bibnamefont{and}
  \bibinfo{author}{\bibfnamefont{E.}~\bibnamefont{Teller}},
  \bibinfo{journal}{J. Chem. Phys.} \textbf{\bibinfo{volume}{21}},
  \bibinfo{pages}{1087} (\bibinfo{year}{1953}).

\bibitem[{\citenamefont{Hastings}(1970)}]{Hastings:1970}
\bibinfo{author}{\bibfnamefont{W.}~\bibnamefont{Hastings}},
  \bibinfo{journal}{Biometrika} \textbf{\bibinfo{volume}{57}},
  \bibinfo{pages}{97} (\bibinfo{year}{1970}).

\bibitem[{\citenamefont{Blanchet}(2006)}]{Blanchet:2006zz}
\bibinfo{author}{\bibfnamefont{L.}~\bibnamefont{Blanchet}},
  \bibinfo{journal}{Living Rev. Rel.} \textbf{\bibinfo{volume}{9}},
  \bibinfo{pages}{4} (\bibinfo{year}{2006}).

\bibitem[{\citenamefont{Ajith et~al.}(2011)\citenamefont{Ajith, Hannam, Husa,
  Chen, Br\"{u}gmann et~al.}}]{Ajith:2009bn}
\bibinfo{author}{\bibfnamefont{P.}~\bibnamefont{Ajith}},
  \bibinfo{author}{\bibfnamefont{M.}~\bibnamefont{Hannam}},
  \bibinfo{author}{\bibfnamefont{S.}~\bibnamefont{Husa}},
  \bibinfo{author}{\bibfnamefont{Y.}~\bibnamefont{Chen}},
  \bibinfo{author}{\bibfnamefont{B.}~\bibnamefont{Br\"{u}gmann}},
  \bibnamefont{et~al.}, \bibinfo{journal}{Phys. Rev. Lett.}
  \textbf{\bibinfo{volume}{106}}, \bibinfo{pages}{241101}
  (\bibinfo{year}{2011}), \eprint{{arXiv:0909.2867 [gr-qc]}}.

\bibitem[{\citenamefont{Buonanno et~al.}(2007)}]{Buonanno:2007pf}
\bibinfo{author}{\bibfnamefont{A.}~\bibnamefont{Buonanno}}
  \bibnamefont{et~al.}, \bibinfo{journal}{Phys. Rev. D}
  \textbf{\bibinfo{volume}{76}}, \bibinfo{pages}{104049}
  (\bibinfo{year}{2007}), \eprint{{arXiv:0706.3732 [gr-qc]}}.

\bibitem[{\citenamefont{Lindblom et~al.}(2008)\citenamefont{Lindblom, Owen, and
  Brown}}]{Lindblom:2008cm}
\bibinfo{author}{\bibfnamefont{L.}~\bibnamefont{Lindblom}},
  \bibinfo{author}{\bibfnamefont{B.~J.} \bibnamefont{Owen}}, \bibnamefont{and}
  \bibinfo{author}{\bibfnamefont{D.~A.} \bibnamefont{Brown}},
  \bibinfo{journal}{Phys. Rev. D} \textbf{\bibinfo{volume}{78}},
  \bibinfo{pages}{124020} (\bibinfo{year}{2008}), \eprint{{arXiv:0809.3844
  [gr-qc]}}.

\bibitem[{PSD(2010{\natexlab{a}})}]{PSDold}
\emph{\bibinfo{title}{{Sensitivity to Gravitational Waves from Compact Binary
  Coalescences Achieved during LIGO's Fifth and Virgo's First Science Run}}}
  (\bibinfo{year}{2010}{\natexlab{a}}), \bibinfo{note}{document
  {LIGO-T0900499-v19} in \url{https://dcc.ligo.org/LIGO-T0900499-v19/public}}.

\bibitem[{PSD(2010{\natexlab{b}})}]{PSD}
\emph{\bibinfo{title}{{Advanced LIGO anticipated sensitivity curves.}}}
  (\bibinfo{year}{2010}{\natexlab{b}}), \bibinfo{note}{document {LIGO-T0900288}
  in \url{https://dcc.ligo.org/LIGO-T0900288-v3/public}}.

\bibitem[{\citenamefont{Aasi et~al.}(2013)}]{S6PE}
\bibinfo{author}{\bibfnamefont{J.}~\bibnamefont{Aasi}} \bibnamefont{et~al.}
  (\bibinfo{collaboration}{LIGO Collaboration, Virgo Collaboration})
  (\bibinfo{year}{2013}), \eprint{arXiv:1304.1775 [gr-qc]}.

\bibitem[{\citenamefont{Allen et~al.}(2003)\citenamefont{Allen, Creighton,
  Flanagan, and Romano}}]{Allen:2002jw}
\bibinfo{author}{\bibfnamefont{B.}~\bibnamefont{Allen}},
  \bibinfo{author}{\bibfnamefont{J.~D.} \bibnamefont{Creighton}},
  \bibinfo{author}{\bibfnamefont{E.~E.} \bibnamefont{Flanagan}},
  \bibnamefont{and} \bibinfo{author}{\bibfnamefont{J.~D.}
  \bibnamefont{Romano}}, \bibinfo{journal}{Phys. Rev. D}
  \textbf{\bibinfo{volume}{67}}, \bibinfo{pages}{122002}
  (\bibinfo{year}{2003}), \eprint{{arXiv}:gr-qc/0205015}.

\bibitem[{\citenamefont{Adams and Cornish}(2010)}]{Adams:2010vc}
\bibinfo{author}{\bibfnamefont{M.~R.} \bibnamefont{Adams}} \bibnamefont{and}
  \bibinfo{author}{\bibfnamefont{N.~J.} \bibnamefont{Cornish}}
  (\bibinfo{year}{2010}), \eprint{{arXiv}:1002.1291}.

\bibitem[{\citenamefont{Rover et~al.}(2011)\citenamefont{Rover, Meyer, and
  Christensen}}]{Rover:2008yp}
\bibinfo{author}{\bibfnamefont{C.}~\bibnamefont{Rover}},
  \bibinfo{author}{\bibfnamefont{R.}~\bibnamefont{Meyer}}, \bibnamefont{and}
  \bibinfo{author}{\bibfnamefont{N.}~\bibnamefont{Christensen}},
  \bibinfo{journal}{Class.Quant.Grav.} \textbf{\bibinfo{volume}{28}},
  \bibinfo{pages}{015010} (\bibinfo{year}{2011}), \eprint{arXiv:0804.3853
  [stat.ME]}.

\bibitem[{\citenamefont{Rover}(2011)}]{Rover:2011qd}
\bibinfo{author}{\bibfnamefont{C.}~\bibnamefont{Rover}},
  \bibinfo{journal}{Phys. Rev. D} \textbf{\bibinfo{volume}{84}},
  \bibinfo{pages}{122004} (\bibinfo{year}{2011}), \eprint{arXiv:1109.0442}.

\bibitem[{\citenamefont{Veitch and Vecchio}(2010)}]{Veitch:2009hd}
\bibinfo{author}{\bibfnamefont{J.}~\bibnamefont{Veitch}} \bibnamefont{and}
  \bibinfo{author}{\bibfnamefont{A.}~\bibnamefont{Vecchio}},
  \bibinfo{journal}{Phys. Rev. D} \textbf{\bibinfo{volume}{81}},
  \bibinfo{pages}{062003} (\bibinfo{year}{2010}), \eprint{arXiv:0911.3820
  [astro-ph.CO]}.

\bibitem[{\citenamefont{Cornish and Littenberg}(2007)}]{Cornish:2007if}
\bibinfo{author}{\bibfnamefont{N.~J.} \bibnamefont{Cornish}} \bibnamefont{and}
  \bibinfo{author}{\bibfnamefont{T.~B.} \bibnamefont{Littenberg}},
  \bibinfo{journal}{Phys. Rev. D} \textbf{\bibinfo{volume}{76}},
  \bibinfo{pages}{083006} (\bibinfo{year}{2007}), \eprint{{arXiv}:0704.1808}.

\bibitem[{\citenamefont{Littenberg and Cornish}(2009)}]{Littenberg:2009bm}
\bibinfo{author}{\bibfnamefont{T.~B.} \bibnamefont{Littenberg}}
  \bibnamefont{and} \bibinfo{author}{\bibfnamefont{N.~J.}
  \bibnamefont{Cornish}}, \bibinfo{journal}{Phys. Rev. D}
  \textbf{\bibinfo{volume}{80}}, \bibinfo{pages}{063007}
  (\bibinfo{year}{2009}), \eprint{{arXiv}:0902.0368}.

\bibitem[{\citenamefont{Littenberg and Cornish}(2010)}]{Littenberg:2010gf}
\bibinfo{author}{\bibfnamefont{T.~B.} \bibnamefont{Littenberg}}
  \bibnamefont{and} \bibinfo{author}{\bibfnamefont{N.~J.}
  \bibnamefont{Cornish}}, \bibinfo{journal}{Phys. Rev. D}
  \textbf{\bibinfo{volume}{82}}, \bibinfo{pages}{103007}
  (\bibinfo{year}{2010}), \eprint{{arXiv}:1008.1577}.

\bibitem[{\citenamefont{Babak et~al.}(2008)}]{Babak:2007zd}
\bibinfo{author}{\bibfnamefont{S.}~\bibnamefont{Babak}} \bibnamefont{et~al.}
  (\bibinfo{collaboration}{Mock LISA Data Challenge Task Force}),
  \bibinfo{journal}{Class. Quantum Grav.} \textbf{\bibinfo{volume}{25}},
  \bibinfo{pages}{114037} (\bibinfo{year}{2008}), \eprint{{arXiv}:0711.2667}.

\bibitem[{\citenamefont{Babak et~al.}(2010)}]{Babak:2009cj}
\bibinfo{author}{\bibfnamefont{S.}~\bibnamefont{Babak}} \bibnamefont{et~al.}
  (\bibinfo{collaboration}{Mock LISA Data Challenge Task Force}),
  \bibinfo{journal}{Class. Quantum Grav.} \textbf{\bibinfo{volume}{27}},
  \bibinfo{pages}{084009} (\bibinfo{year}{2010}), \eprint{{arXiv}:0912.0548}.

\bibitem[{\citenamefont{Abbott et~al.}(2008)}]{Abbott:2007rh}
\bibinfo{author}{\bibfnamefont{B.}~\bibnamefont{Abbott}} \bibnamefont{et~al.}
  (\bibinfo{collaboration}{LIGO Scientific Collaboration}),
  \bibinfo{journal}{Astrophys. J.} \textbf{\bibinfo{volume}{681}},
  \bibinfo{pages}{1419} (\bibinfo{year}{2008}), \eprint{{arXiv}:0711.1163}.

\bibitem[{\citenamefont{{Abbott} et~al.}(2008)}]{2008ApJ...683L..45A}
\bibinfo{author}{\bibfnamefont{B.}~\bibnamefont{{Abbott}}} \bibnamefont{et~al.}
  (\bibinfo{collaboration}{LIGO Scientific Collaboration}),
  \bibinfo{journal}{Astrophys. J. Lett.} \textbf{\bibinfo{volume}{683}},
  \bibinfo{pages}{L45} (\bibinfo{year}{2008}), \eprint{0805.4758}.

\bibitem[{\citenamefont{Abbott et~al.}(2009)}]{Abbott:2009up}
\bibinfo{author}{\bibfnamefont{B.}~\bibnamefont{Abbott}} \bibnamefont{et~al.}
  (\bibinfo{collaboration}{LIGO Scientific Collaboration}),
  \bibinfo{journal}{Phys. Rev. D} \textbf{\bibinfo{volume}{80}},
  \bibinfo{pages}{102002} (\bibinfo{year}{2009}), \eprint{{arXiv:0904.4910
  [gr-qc]}}.

\bibitem[{\citenamefont{{Abbott} et~al.}(2009)}]{StochasticBackgroundS5}
\bibinfo{author}{\bibfnamefont{B.}~\bibnamefont{{Abbott}}} \bibnamefont{et~al.}
  (\bibinfo{collaboration}{LIGO Scientific Collaboration, The Virgo
  Collaboration}), \bibinfo{journal}{Nature} \textbf{\bibinfo{volume}{460}},
  \bibinfo{pages}{990} (\bibinfo{year}{2009}), \eprint{{arXiv}:0910.5772
  [astro-ph.CO]}.

\bibitem[{\citenamefont{Swendsen and Wang}(1986)}]{Swendsen:1986}
\bibinfo{author}{\bibfnamefont{R.}~\bibnamefont{Swendsen}} \bibnamefont{and}
  \bibinfo{author}{\bibfnamefont{J.}~\bibnamefont{Wang}},
  \bibinfo{journal}{Phys. Rev. Lett.} \textbf{\bibinfo{volume}{57}},
  \bibinfo{pages}{2607} (\bibinfo{year}{1986}).

\bibitem[{\citenamefont{Goggans and Chi}(2004)}]{Goggans:2004}
\bibinfo{author}{\bibfnamefont{P.}~\bibnamefont{Goggans}} \bibnamefont{and}
  \bibinfo{author}{\bibfnamefont{Y.}~\bibnamefont{Chi}},
  \emph{\bibinfo{title}{Bayesian Inference and Methods in Science and
  Engineering}} (\bibinfo{publisher}{American Institute of Physics},
  \bibinfo{address}{USA}, \bibinfo{year}{2004}).

\bibitem[{\citenamefont{Rover et~al.}(2007)\citenamefont{Rover, Meyer, and
  Christensen}}]{Rover:2006bb}
\bibinfo{author}{\bibfnamefont{C.}~\bibnamefont{Rover}},
  \bibinfo{author}{\bibfnamefont{R.}~\bibnamefont{Meyer}}, \bibnamefont{and}
  \bibinfo{author}{\bibfnamefont{N.}~\bibnamefont{Christensen}},
  \bibinfo{journal}{Phys. Rev. D} \textbf{\bibinfo{volume}{75}},
  \bibinfo{pages}{062004} (\bibinfo{year}{2007}), \eprint{arXiv:gr-qc/0609131}.

\bibitem[{\citenamefont{Nissanke et~al.}(2011)\citenamefont{Nissanke, Sievers,
  Dalal, and Holz}}]{Nissanke:2011ax}
\bibinfo{author}{\bibfnamefont{S.}~\bibnamefont{Nissanke}},
  \bibinfo{author}{\bibfnamefont{J.}~\bibnamefont{Sievers}},
  \bibinfo{author}{\bibfnamefont{N.}~\bibnamefont{Dalal}}, \bibnamefont{and}
  \bibinfo{author}{\bibfnamefont{D.}~\bibnamefont{Holz}},
  \bibinfo{journal}{Astrophys.J.} \textbf{\bibinfo{volume}{739}},
  \bibinfo{pages}{99} (\bibinfo{year}{2011}), \eprint{arXiv:1105.3184
  [astro-ph]}.

\bibitem[{\citenamefont{Veitch et~al.}(2012)\citenamefont{Veitch, Mandel,
  Aylott, Farr, Raymond et~al.}}]{Veitch:2012df}
\bibinfo{author}{\bibfnamefont{J.}~\bibnamefont{Veitch}},
  \bibinfo{author}{\bibfnamefont{I.}~\bibnamefont{Mandel}},
  \bibinfo{author}{\bibfnamefont{B.}~\bibnamefont{Aylott}},
  \bibinfo{author}{\bibfnamefont{B.}~\bibnamefont{Farr}},
  \bibinfo{author}{\bibfnamefont{V.}~\bibnamefont{Raymond}},
  \bibnamefont{et~al.}, \bibinfo{journal}{Phys. Rev. D}
  \textbf{\bibinfo{volume}{85}}, \bibinfo{pages}{104045}
  (\bibinfo{year}{2012}), \eprint{{arXiv:1201.1195 [astro-ph.HE]}}.

\bibitem[{S5D(2010)}]{S5DataQuality}
\emph{\bibinfo{title}{{Data Quality and Veto Choices of S5 Lowmass CBC
  Searches}}} (\bibinfo{year}{2010}), \bibinfo{note}{document
  {LIGO-T1000056-v5} in \url{https://dcc.ligo.org/LIGO-T1000056-v5/public}}.

\bibitem[{\citenamefont{Buonanno et~al.}(2009)\citenamefont{Buonanno, Iyer,
  Ochsner, Pan, and Sathyaprakash}}]{Buonanno:2009zt}
\bibinfo{author}{\bibfnamefont{A.}~\bibnamefont{Buonanno}},
  \bibinfo{author}{\bibfnamefont{B.}~\bibnamefont{Iyer}},
  \bibinfo{author}{\bibfnamefont{E.}~\bibnamefont{Ochsner}},
  \bibinfo{author}{\bibfnamefont{Y.}~\bibnamefont{Pan}}, \bibnamefont{and}
  \bibinfo{author}{\bibfnamefont{B.}~\bibnamefont{Sathyaprakash}},
  \bibinfo{journal}{Phys Rev D} \textbf{\bibinfo{volume}{80}},
  \bibinfo{pages}{084043} (\bibinfo{year}{2009}),
  \eprint{{arXiv:gr-qc/0907.0700}}.

\end{thebibliography}

\end{document}